\documentstyle[12pt,bezier]{article}
\begin{document}
% ***************    NEW COMMANDS   *******************
\def \inbar{\vrule height1.5ex width.4pt depth0pt}
\def \xC{\relax\hbox{\kern.25em$\inbar\kern-.3em{\rm C}$}}
\def \xR{\relax{\rm I\kern-.18em R}}
\newcommand{\xZ}{Z \hspace{-.08in}Z}
\newcommand{\xbe}{\begin{equation}}
\newcommand{\xee}{\end{equation}}
\newcommand{\xbea}{\begin{eqnarray}}
\newcommand{\xeea}{\end{eqnarray}}
\newcommand{\xnn}{\nonumber}
\newcommand{\xkt}{\rangle}
\newcommand{\xbr}{\langle}
\newcommand{\uuu}{{\mbox{\Large$u$}}}
\newcommand{\xcun}{{\mbox{\footnotesize${\cal N}$}}}
\title{Non-Abelian Geometric Phase for General
Three-Dimensional Quantum Systems}
\author{Ali Mostafazadeh\thanks{E-mail: 
alimos@phys.ualberta.ca}\\ \\
Theoretical Physics Institute, University of Alberta, \\
Edmonton, Alberta,  Canada T6G 2J1.}
\date{August 1996}
\maketitle

\begin{abstract}
Adiabatic $U(2)$ geometric phases are studied for arbitrary
quantum systems with a three-dimensional Hilbert space.
Necessary and sufficient conditions for the occurrence of the
non-Abelian geometrical phases are obtained without actually
solving the full eigenvalue problem for the instantaneous Hamiltonian. 
The parameter space of such systems which has
the structure of $\xC P^2$ is explicitly constructed. 
The results of this article are applicable for arbitrary multipole
interaction Hamiltonians $H=Q^{i_1,\cdots i_n}J_{i_1}\cdots
J_{i_n}$ and their linear combinations for spin $j=1$ systems.
In  particular it is shown that the nuclear quadrupole
Hamiltonian $H=Q^{ij}J_iJ_j$ does actually lead to
non-Abelian geometric phases for $j=1$.  This system, being
bosonic, is time-reversal-invariant. Therefore it cannot
support  Abelian adiabatic geometrical phases. 
\end{abstract}

\baselineskip=18pt

\newpage

\section{Introduction}

In 1984 Berry published a beautiful article \cite{berry1984}, in
which he systematically studied what is now called the Berry
phase or the adiabatic geometrical phase.\footnote{Manifestations
of the phenomenon of  the geometric phase have been known
to chemists \cite{mead} long before Berry's article.} Berry's
observation has since attracted the attention of a large number
of theoretical and experimental physicists. One of the most
important developments in the subject has been the discovery
of the non-Abelian analogues of the adiabatic geometrical
phase by Wilczek and Zee \cite{wi-ze}. This development has
unraveled some interesting manifestations of the non-Abelian
gauge theories in non-relativistic quantum mechanics,
particularly in the realm of molecular physics \cite{bohm-qm}. 

Perhaps one of the most important results which contributed
to a better understanding of the Abelian and  non-Abelian
geometrical phases is Simon's identification of Berry's Abelian
phase with a holonomy element of a $U(1)$ spectral bundle
over the space of the environmental parameters of the system,
\cite{simon}. In the language of fiber bundles the non-Abelian
phases of Wilczek and Zee correspond to holonomy elements
of the $U(\xcun)$ spectral bundle associated with an
$\xcun$-fold degenerate eigenvalue of the Hamiltonian.

The range of the environmental parameters $R$, i.e., the
structure of the parameter space $M$ whose points are
coordinatized by $R=(R^1,\cdots,R^n)$, is determined
by the condition of the stability of the degeneracy structure
of the Hamiltonian, i.e., by the condition that
during any possible evolution of the parameters --- for any
smooth curve $C:[0,T]\to M$ --- the degeneracy structure
of the instantaneous Hamiltonian $H(t)=H[C(t)]$, must
not change. Here one also assumes that the eigenvalues
$E_n[R]$ and eigenvectors $|n;R\xkt$ of the Hamiltonian
$H[R]$ are smooth functions of the parameters $R\in M$.
Therefore, different degeneracy structures of the
Hamiltonian correspond to distinct parameter spaces.

Although, there are by now hundreds of publications
on geometric phases, the number of the specific examples
which have been worked out in detail is quite few. The
best-known examples which lead to Abelian geometric
phases are
	\begin{itemize}
	\item[1)] Berry's original example of a magnetic dipole
	(a spin) in a rotating magnetic field with the Hamiltonian:
	\xbe
	H[R]=b~\sum_{i=1}^3R^iJ_i\;,
	\label{h1}
	\xee
where $b$ is a constant (Larmor frequency), $R^i$ are the
Cartesian coordinates of the tip of the magnetic field and $J_i$
are angular momentum operators, i.e., generators
of the dynamical group $SU(2)$, \cite{berry1984,bohm-qm}.
The parameter space of this system is the two-dimensional
sphere $S^2=SU(2)/U(1)$;
	\item[2)] the generalized harmonic oscillator 
\cite{hannay,berry1985}, whose Hamiltonian can also
be put in the form (\ref{h1}) by identifying $J_i$ with the
generators of $SU(1,1)$, \cite{jackiw}. The parameter
space of this system is the hyperbolic sphere $SU(1,1)
/U(1)$.
	\end{itemize}
Immediate generalizations of these examples are obtained
by taking an arbitrary dynamical group $G$ and requiring the
Hamiltonian $H$ to belong to a unitary irreducible representation
of the Lie algebra  ${\cal G}$ of $G$, i.e., choosing $H$ to be a
linear combination of  (the representation of) the generators
of $G$. For a compact semisimple group $G$, I have shown in
Ref.~\cite{p7}, that the relevant parameter space is in general
a subspace of the flag manifold $G/T$, where $T$ is a maximal
torus (largest Abelian subgroup) of $G$. This result holds for
arbitrary quantum systems whose Hilbert space is
finite-dimensional. For in this case the Hamiltonian is in general
a finite-dimensional Hermitian matrix.  Therefore it belongs to
the Lie algebra $u(N)$ of the group $U(N)$, where $N$ is the
dimension of the Hilbert space.

For the non-Abelian phase, the known examples are
	\begin{itemize}
	\item[3)] the original  example of Wilczek and Zee
\cite{wi-ze}, which involves an $(\xcun +1)$-dimensional
system with an $\xcun$-fold degeneracy. The Hamiltonian
of this system is obtained by similarity transformations
(rotations) by elements of $SO(\xcun+1)$ of a diagonal $(\xcun+1)
\times (\xcun+1)$ matrix which has only one non-zero entry.
Clearly, the system has an $SO(\xcun)$ symmetry by
construction.  Hence the parameter space is
$SO(\xcun +1)/SO(\xcun)=S^\xcun$.
	\item[4)] the fermionic (half  odd integer spin) systems
associated with nuclear quadrupole Hamiltonians of the form:
	\xbe
	H=Q^{ij}J_iJ_j\;.
	\label{h2}
	\xee
These systems which were first studied in the context of
geometric phase by Mead \cite{mead2} and subsequently
by Avron, et al \cite{av1,av2}, involve Kramers degeneracy.
The parameter space is $S^4$. As it is shown in \cite{av2},
for the bosonic case (integer spin) the Berry connection
one-form is exact \cite{nakahara} (the Berry gauge
potential is pure gauge). Therefore, the (Abelian) Berry phase
angle vanishes. This is a direct consequence of the fact that the
bosonic systems of this type are time-reversal-invariant. The
same conclusion cannot be reached for the non-Abelian case,
however. The simplest nontrivial case is for spin $j=1$, where
the Hilbert space is three-dimensional. This problem turns
out to be also relevant to the manifestations of the geometric
phase for relativistic scalar fields in Bianchi type IX
cosmological backgrounds \cite{p19}. 
	\end{itemize}

The study of the geometrical phase for
three and higher-dimensional Hilbert spaces
is clearly plagued with the difficulties involving the
solution of the eigenvalue problem for the instantaneous
Hamiltonian. Even in the three-dimensional case, the
characteristic polynomial is of order three which makes
the direct solution of the eigenvalue problem quite
complicated. The motivation for the present article has
been the simple observation that the existence of a
(two-fold) degeneracy can be exploited to identify the
appropriate parameter space over which one can
obtain the eigenvalues and the eigenvectors of the
Hamiltonian straightforwardly. Unlike the direct approach,
in which one tries to solve the full eigenvalue
problem for the Hamiltonian and then obtain the
parameter space by equating two of the eigenvalues,
this method only involves the solution of simple quadratic
equations. This is demonstrated in section~2. This section
also includes a detailed discussion of the parameter
space. Section~3 includes the computation of the
geometric phases. The general results are then applied
to the quadrupole and multipole Hamiltonians (\ref{h2}) in
section~4. 

In the remainder of this section, I briefly review the
non-Abelian geometrical phase of Wilzcek and Zee.

Berry's investigation of the adiabatic geometrical
phase uses the quantum adiabatic theorem \cite{schiff}.
If the time-dependence of the Hamiltonian $H(t)=H[R(t)]$
justifies the validity of the adiabatic approximation,
an initial state vector which is an eigenvector
of the initial Hamiltonian $H(0)=H[R(0)]$, evolves
according to the Schr\"odinger equation, in such a
way that it always remains an eigenvector of the
instantaneous Hamiltonian $H(t)=H[R(t)]$. If the
evolving state vector corresponds to an $\xcun$-fold
degenerate eigenvalue $E_n[R(t)]$, then the adiabatic
theorem states that it must always belong to the
$\xcun$-dimensional degeneracy subspace
${\cal H}_n[R(t)]$ associated with $E_n[R(t)]$.
If the Hamiltonian depends periodically on time,
i.e., the curve $C:[0,T]\to M$ is closed, then after
a period, the Hamiltonian, its eigenvalues, and
the corresponding degeneracy subspaces return
to their original form, i.e., $H(T)=H[R(T)]=H[R(0)]=H(0)$,
$E_n(T)=E_n[R(T)]=E_n[R(0)]=E_n(0)$, and
${\cal H}_n(T)={\cal H}_n[R(T)]={\cal H}_n[R(0)]=
{\cal H}_n(0)$. Therefore the evolving state vector
$|\psi(T)\xkt$ belongs to the same degeneracy
subspace as the initial state vector $|\psi(0)\xkt$.
Since the evolution is supposed to be
unitary, there exists a $U(\xcun)$ matrix relating
$|\psi(T)\xkt$ and $|\psi(0)\xkt$, which is given by
\cite{wi-ze}:
	\xbe
	e^{-\frac{i}{\hbar}\int_0^T E_n(t)}\:
	{\cal P}\left[ e^{i\oint_C A_n}\right]\;.
	\label{ge-ph}
	\xee
Here ${\cal P}$ is the path-ordering operator
\cite{nakahara} and $A_n$ is a $u(\xcun)$-valued
(connection) one-form whose matrix elements
are locally given by:
	\xbe
	A_n^{ab}=i\xbr n,a;R|\frac{\partial}{\partial R^i}
	|n,b;R\xkt\: dR^i=i\xbr n,a;R|d|n,b;R\xkt\;.
	\label{connection}
	\xee
In this equation $a$ and $b$ are degeneracy labels,
$\{ |n,a;R\xkt\}$, with $a=1,\cdots,\xcun$, is an
orthonormal local basis of ${\cal H}_n[R]$,
and $d$ denotes the exterior derivative operator
on $M$.

The first exponential in (\ref{ge-ph}) is called
the {\em dynamical phase}, whereas the second
(path-ordered) exponential is called the
{\em non-Abelian adiabatic geometrical phase}. As it
is seen  from (\ref{ge-ph}), unlike the dynamical
phase, the geometrical phase only depends on
the shape of the curve $C$ in $M$ and not on its
parameterization. Usually this property of the
geometric phase is offered as a justification of its
geometrical nature. This point of view does not
however do justice to the most intriguing geometrical
properties of this phase which are best described in
terms of the geometry of spectral bundles on the space of
parameters and the universal classifying bundles on the
projective Hilbert space. For a thorough discussion of
the mathematical structure of the geometric phase see
Refs.~\cite{p6,p7} and references therein.

Finally let me emphasize that the geometrical
phase is not a topological quantity in general. By definition
a topological phase, such as the Aharonov-Bohm
phase, is invariant under smooth deformations of
the curve $C$. This is not generally the case for 
arbitrary geometrical phases. Topological phases
of this type form a proper subset of all geometrical
phases. There have been some arguments in the
literature concerning the topological content of
the geometrical phase \cite{kr} (see also
\cite{jackiw} and \cite{gi}) in which the removal of
a geometrical phase via a smooth deformation of
the functional dependence of the Hamiltonian on the
parameters, or  through a time-dependent canonical
transformation, has been used to justify the
attribution of the term `trivial' to these phases. A
typical example of this type of removable geometrical
phases occurs for the generalized harmonic oscillator.
For a specific physical system, however, such
deformations or canonical transformations cannot
be freely affected. In this sense topologically trivial
geometrical phases such as those associated with
the generalized harmonic oscillator are as physically
significant as their topologically nontrivial
counterparts. Therefore, in this article, I shall
use the phrase `trivial geometrical phase' to
mean that the corresponding matrix-valued `phase
angle' vanishes, i.e., the geometrical phase does
not exist. A precise characterization of
the topological content of the Abelian adiabatic
geometrical phases which are associated with compact
semisimple dynamical groups (finite-dimensional
Hilbert spaces) is given in Ref.~\cite{p7}.

\section{Three-Dimensional Systems}
The simplest non-Abelian geometric phases
belong to $U(2)$. Hence the first nontrivial case
which leads to non-Abelian $U(2)$-geometric phases
is when the dimension of the Hilbert space is three.
In this cases, the Hamiltonian is a $3\times 3$
Hermitian matrix which can be viewed as
an element of the Lie algebra $u(3)$, i.e., 
	\xbe
	H[R]=b \sum_{i=0}^8 R^i\lambda_i\;,
	\label{h3}
	\xee
where $R^i$ are real parameters and $\lambda_i$ are 
generators of U(3) in the defining representation. For
example one can take $\lambda_0$ to be the identity
matrix $I_{3\times 3}$, and identify $\lambda_1,
\cdots,\lambda_8$ with the Gell-Mann matrices \cite{gri}.
In the remainder of this article a unit system is used
where $b=1$. The adiabaticity assumption may then
be expressed as $T\gg 1$.

It is shown in Ref.~\cite{p7}, that as a consequence of
some group theoretical considerations the parameter
space $M$ is in general (a subspace of) the manifold 
$SU(3)/U(2)=\xC P^2$. In this section, I shall explore
the parameter space for the case where one of the
eigenvalues of the Hamiltonian is doubly degenerate.
For this particular case it is quite straightforward to
see that indeed $M=\xC P^2$.
Let us first express the Hamiltonian $H$ in the form:
	\xbe
	H[R]=\uuu [R]\: H_D\:\uuu^{-1}[R]\;,
	\label{similarity}
	\xee
where $\uuu[R]\in U(3)$ and $H_D$ is diagonal:
	\xbe
	H_D=\left(\begin{array}{ccc}
	E_1&0&0\\
	0&E_2&0\\
	0&0&E_2\end{array}\right)\;.
	\label{h-diag}
	\xee
In $U(3)$ there are two distinct  $U(1)$ and $U(2)$
subgroups which respectively leave the degeneracy
subspaces ${\cal H}_1$ and ${\cal H}_2$ invariant.
Hence the true parameter space is $U(3)/[U(2)\times
U(1)]=\xC P^2$. This argument
does not however provide a concrete characterization
of the parameter space unless one actually solves the
full eigenvalue equation for the Hamiltonian, i.e., finds
the explicit expression for $\uuu[R]$ in terms of $R^i$.
The main result of this article is the fact that
due to the presence of the degeneracy this procedure
can be replaced with another much more manageable
method which leads to an explicit construction of
the parameter space and a direct computation of the
associated geometric phases.

In order to implement the condition on the degeneracy,
let me write the Hamiltonian in the form:
	\xbe
	H=\left(\begin{array}{ccc}
	r&\xi^*&\zeta^*\\
	\xi&s&\kappa^*\\
	\zeta&\kappa&t
	\end{array}\right)\;,
	\label{h4}
	\xee
where $r,~s,~t$ are real and $\xi,~\zeta,~
\kappa$ are complex parameters related to $R^i$,
according to:
	\xbea
	r&=&R^0+R^3+R^8/\sqrt{3}\;,~~~
	s\:=\: R^0-R^3+R^8/\sqrt{3}\;,~~~
	t\:=\:R^0-2R^8/\sqrt{3}\;,\xnn\\
	\xi&=&R^1+iR^2\;,~~~
	\zeta\:=\:R^4+iR^5\;,~~~
	\kappa\:=\:R^6+iR^7\;.
	\label{=R}
	\xeea
These equations are obtained  using the expression
for the Gell-Mann matrices, as given for
example in Ref.~\cite{gri}, in Eq.~(\ref{h3}) and
comparing this equation with (\ref{h4}).

Furthermore, since the addition of a multiple of
the identity operator does not have any physical
implications one can alternatively consider the
Hamiltonian
	\xbe
	H':=H-E_2 I_{3\times 3}=\left(\begin{array}{ccc}
	r'&\xi^*&\zeta^*\\
	\xi&s'&\kappa^*\\
	\zeta&\kappa&t'
	\end{array}\right)\;,
	\label{h'4}
	\xee
where 
	\xbe
	r'=r-E_2\;,~~~~~ s'=s-E_2\;,~~~~~t'=t-E_2\;.
	\label{primes=}
	\xee
Clearly $H$ and $H'$ have identical eigenvectors.
Their eigenvalues are related by $E'_1=E_1-E_2$ and
$E'_2=0$, where $E'_2$ corresponds to the
degenerate eigenvalue. Clearly, it is $E'_1$ and the
common eigenvectors which are physically
significant.\footnote{Strictly speaking the sign
of $E'_1$ is also a conventional choice.}

Next consider the eigenvalue problem for the
Hamiltonian $H'$. The corresponding characteristic
polynomial is given by:
	\xbea
	P(E')&:=&\det [H'-E' I_{3\times 3}]\xnn\\
	&=&-E^{'3}+(r'+s'+t')E^{'2}+
	(-r's'-s't'-t'r' +|\xi|^2+|\zeta|^2+|\kappa|^2)E'+\xnn\\
	&& r's't'+\kappa\xi\zeta^*+\kappa^*\xi^*\zeta-
	r'|\kappa|^2-s'|\zeta|^2-t'|\xi|^2\;.
	\label{ch-po}
	\xeea
On the other hand, since eigenvalues $E'_1$ and
$E'_2$ are the roots of $P(E')$ and $E'_2=0$ is doubly
degenerate, 
	\xbe
	P(E')=-(E'-E'_1)E^{'2}\;.
	\label{ch-po-2}
	\xee
Comparing the two expressions (\ref{ch-po}) and
(\ref{ch-po-2}) one finds:
	\xbea
	&&E'_1=r'+s'+t' \;,
	\label{e1}\\
	&&-r's'-s't'-t'r' +|\xi|^2+|\zeta|^2+|\kappa|^2=0\;,
	\label{e2}\\
	 &&r's't'+\kappa\xi\zeta^*+\kappa^*\xi^*\zeta-
	r'|\kappa|^2-s'|\zeta|^2-t'|\xi|^2=0\;.
	\label{e3}
	\xeea

Furthermore, the fact that $E'_2=0$ is doubly degenerate
implies that the rows of the matrix $H'-E'_2I_{3\times 3}=H'$
must be mutually linearly dependent. Equivalently the
cofactors of the matrix elements of $H'$ must vanish. This
leads to:
	\xbea
	s't'-|\kappa|^2&=&0\;,
	\label{c1}\\
	t'\xi-\zeta\kappa^*&=&0\;,
	\label{c2}\\
	s'\zeta-\xi\kappa&=&0\;,
	\label{c3}\\
	r't'-|\zeta|^2&=&0\;,
	\label{c4}\\
	r'\kappa-\xi^*\zeta&=&0\;,
	\label{c5}\\
	r's'-|\xi|^2&=&0\;.
	\label{c6}
	\xeea
In view of these equations, Eqs.~(\ref{e2}) and (\ref{e3})
are satisfied automatically. Moreover, either at most two of
the parameters $r'$, $s'$ and $t'$ vanish or all of them
have the same sign as $E'_1$. Note that, as a result of 
Eq.~(\ref{e1}) and the non-degeneracy requirement
$E'_1\neq 0$, $r'$, $s'$ and $t'$ cannot vanish
simultaneously.

Eqs.~(\ref{c1})-(\ref{c6}) are indeed not independent.
They can be reduced to the following four equations:
	\xbea
	\xi=\sqrt{r's'}\:e^{i\gamma}\;,&&
	\label{ct6}\\
	\zeta=\sqrt{r't'}\:e^{i\eta}\;,&&
	\label{ct4}\\
	\kappa=\sqrt{s't'}\:e^{i\theta}\;,&&
	\label{ct1}\\
	r's't'[e^{i\eta}-e^{i(\gamma+\theta)}]=0\;.&&
	\label{ct2}
	\xeea
Thus there are five independent real parameters. In addition,
one knows that a rescaling of the Hamiltonian by a non-zero
real function of the variables leaves the eigenvectors
unchanged. Hence as far as the geometric phases are
concerned, one may reduce the number of real parameters
to four. 

In the calculation of the geometric phases, I shall not
explicitly perform this reduction. Hence the results will
be valid for arbitrary $3\times 3$ Hamiltonians. The final
expressions for the geometric phases (connection
one-forms) however are expected to be invariant
under the simultaneous scaling of the matrix elements
of the Hamiltonian. In view of Eqs.~(\ref{ct6})-(\ref{ct1}),
this means that they must only involve the ratios of  the
parameters $r',~s'$ and $t'$. 

Therefore the true parameters of this system are
the ratios of $r',~s'$ and $t'$, and the angles $\gamma$
and $\theta$. In fact it is not difficult to show that these
parameters yield a local coordinate representation of
$\xC P^2$.

To see this let me recall the homogeneous coordinates
on $\xC P^2$, \cite{nakahara}:
	\xbe
	(z^1,z^2,z^3)\equiv (1,\frac{z^2}{z^1},
	\frac{z^3}{z^1})\equiv(\frac{z^2}{z^1},
	\frac{z^3}{z^1})=:(\rho_{21}\:e^{i\phi_{21}},
	\rho_{31}\:e^{i\phi_{31}})\;,
	\label{homo-coor}
	\xee
corresponding to the local patch $O_1\subset
\xC P^2$ defined by $z^1\neq 0$. In (\ref{homo-coor}),
$z^\mu\in\xC$, $\rho_{\mu1}:=|z^\mu/z^1|$, and
$\phi_{\mu1}:=|z^1| z^\mu/(|z^\mu|z^1)$, with $\mu=
1,2,3$. 	

The analogy with the parameters of the doubly
degenerate Hamiltonian $H'$ can then be
expressed by $z^1=\xi$, $z^2=\zeta$,
and $z^3=\kappa$. In terms of the real variables
one has 
	\xbe
	\rho_{21}=\sqrt{\frac{t'}{s'}}\;,~~~~
	\rho_{31}=\sqrt{\frac{t'}{r'}}\;,~~~~
	\phi_{21}=-(\theta+2\gamma)\;,~~~~
	\phi_{31}=\theta-\gamma\;,
	\label{real-coor}
	\xee
where the patch $O_1$ is defined by
$r'\neq 0,~s'\neq 0$. Similar relations hold for
the other two patches $O_2:~z^2=\zeta\neq 0~(
r'\neq 0,~s'\neq 0)$ and $O_3:~z^3=\kappa\neq
0~(s'\neq 0,t'\neq 0)$. This leaves only the
case where two of the parameters $r',s'$ and $t'$
vanish. In this case, the Hamiltonian $H'$ is already
diagonal and the eigenvectors are $(1,0,0)$,
$(0,1,0)$ and  $(0,0,1)$. Therefore the connection
one-forms (\ref{connection}) vanish identically.
This case can then be excluded from the
true parameter space. This concludes the explicit
construction of the parameter space $M=\xC P^2$.

Next let me express the necessary and sufficient
conditions for the existence of a doubly degenerate
eigenvalue in terms of the parameters of the
original Hamiltonian $H$. Clearly the case $\xi=\zeta=
\kappa=0$ is trivial. This leaves one with five distinct cases. Namely,
	\begin{itemize}
	\item[1)] \underline{$\xi\neq 0$, $\zeta\neq 0$ and $\kappa
\neq 0$:}\\
In this case Eqs.~(\ref{ct6})-(\ref{ct2}) may be used to
show:
	\xbea
	\frac{\xi\kappa}{\zeta}&\in&\xR^+\;,
	\label{condi-c1-1}\\
	r'&=&\pm\left| \frac{\xi\zeta}{\kappa}\right|\;,~~~~~
	s'\:=\:\pm\frac{\xi\kappa}{\zeta}\;,~~~~~
	t'\:=\:\pm\left| \frac{\zeta\kappa}{\xi}\right|\;.
	\label{primes=2}
	\xeea
Eqs.~(\ref{primes=}) and (\ref{primes=2}), then lead
to
	\xbe
	E_n=r\mp\left| \frac{\xi\zeta}{\kappa}\right|=
	s\mp\frac{\xi\kappa}{\zeta}=
	t\mp\left| \frac{\zeta\kappa}{\xi}\right|\;.
	\label{condi-c1-2}
	\xee
The last two equations together with (\ref{condi-c1-1})
are equivalent to Eqs.~(\ref{ct6})-(\ref{ct2}). They serve
as the necessary and sufficient condition for the
double degeneracy of $E_2$. More symmetric
expressions for $E_1$ and $E_2$ are:
	\xbea
	E_1&=&\frac{1}{3}\:\left[r+s+t\pm 
	2(\left| \frac{\xi\zeta}{\kappa}\right|+
	\frac{\xi\kappa}{\zeta}+\left| \frac{\zeta\kappa}{
	\xi}\right|)\right]\;,\label{E1=}\\
	E_2&=&\frac{1}{3}\:\left[r+s+t\mp 
	\left| \frac{\xi\zeta}{\kappa}\right|+
	\frac{\xi\kappa}{\zeta}+\left| \frac{\zeta\kappa}{
	\xi}\right|~\right]\;.\label{E2=}
	\xeea
Note also that in Eqs.~(\ref{primes=2})-(\ref{E2=})
either the top or the bottom sign must be chosen.
Both choices are physically equivalent.
	\item[2)] \underline{$\xi=\zeta=0$ and $\kappa\neq 0$:}\\
In this case $r'=0$. Hence
	\xbe
	E_2=r\;,~~~~~s'=s-r\;,~~~~~t'=t-r\;,~~~~~
	E_1=-r+s+t\;.
	\label{E1-E2-c2}
	\xee
Furthermore, the necessary and sufficient condition
for the double degeneracy of $E_2$ is
	\xbe
	|\kappa|^2=(s-r)(t-r)\;.
	\label{condi-c2}
	\xee
	\item[3)] \underline{$\xi=\kappa=0$ and $\zeta\neq 0$:}\\
In this case $s'=0$ and one has:
	\xbe
	E_2=s\;,~~~~~r'=r-s\;,~~~~~t'=t-s\;,~~~~~
	E_1=r-s+t\;.
	\label{E1-E2-c3}
	\xee
The necessary and sufficient condition for the
occurrence of  a double degeneracy is
	\xbe
	|\zeta|^2=(r-s)(t-s)\;.
	\label{condi-c4}
	\xee
	\item[4)] \underline{$\zeta=\kappa=0$ and $\xi\neq 0$:}\\
In this case $t'=0$ and one has:
	\xbe
	E_2=t\;,~~~~~r'=r-t\;,~~~~~s'=s-t\;,~~~~~
	E_1=r+s-t\;.
	\label{E1-E2-c4}
	\xee
The necessary and sufficient condition for the
occurrence of  a double degeneracy is
	\xbe
	|\xi|^2=(r-t)(s-t)\;.
	\label{condi-c3}
	\xee
	\item[5)] \underline{only one of  the parameters 
$\xi,\zeta,\kappa$ is zero:}\\
In this case, eigenvalues of H cannot be doubly
degenerate.
\end{itemize}

\section{Connection One-Forms}

In order to compute the connection one-forms associated
with the eigenvalues of the Hamiltonian, one must consider
the first four cases of the above list separately. 
In terms of the local coordinate patches $O_\mu$ of the
parameter space $M=\xC P^2$, these cases correspond to
	\begin{itemize}
	\item[1)]  the intersection $O_1\cap O_2\cap O_3$
	in which $r'\neq 0$, $s'\neq 0$ and $t'\neq 0$;
	\item[2)] the subset $O_3- O_1\cap O_2\cap
	O_3$, in which $r'=0$, $s'\neq 0$ and $t'\neq 0$;
	\item[3)] the subset $O_2- O_1\cap O_2\cap
	O_3$, in which $s'=0$, $r'\neq 0$ and $t'\neq 0$;
	\item[4)] the subset $O_1- O_1\cap O_2\cap
	O_3$, in which $t'=0$, $r'\neq 0$ and $s'\neq 0$,
	\end{itemize}
respectively.

The computation of the connection one-forms for all these
four cases involves using Eqs.~(\ref{ct6})-(\ref{ct2}) to
obtain the eigenvectors of the Hamiltonian $H'$. This
is indeed quite straightforward. Having obtained the
eigenvectors, one then computes the connection
one-forms using Eq.~(\ref{connection}).

\subsection*{Case 1: $r'\neq 0,~s'\neq 0,~t'\neq 0$}
The eigenvectors are given by:
	\xbe
	|1\xkt=N_1\:\left(\begin{array}{c}
	\sqrt{r'}\:e^{-i\gamma}\\
	\sqrt{s'}\\
	\sqrt{t'}\:e^{i\theta}
	\end{array}\right)\;,~~~~~
	|2,1\xkt=N_2\:\left(\begin{array}{c}
	-\sqrt{s'}\:e^{-i\gamma}\\
	\sqrt{r'}\\0
	\end{array}\right)\;,~~~~~
	|2,2\xkt=N_1N_2\:\left(\begin{array}{c}
	\sqrt{r't'}\:e^{-i\gamma}\\
	\sqrt{s't'}\\
	-(r'+s')\:e^{i\theta}
	\end{array}\right)\;,
	\label{eq-ve-c1}
	\xee
where $N_1:=(r'+s'+t')^{-1/2}=E_1^{'-1/2}$ and $N_2:=
(r'+s')^{-1/2}$.

Substituting these equations in (\ref{connection}), and
performing the necessary algebra one finds:
	\xbea
	A_1&=&\frac{d\gamma}{1+\frac{s'+t'}{r'}}+
	\frac{d\theta}{1+\frac{t'}{r'+s'}}\;,
	\label{c1-a1}\\
	A_2&=&\mbox{\Large$\left(\begin{array}{cc}
	\frac{d\gamma}{1+\frac{r'}{s'}}&
	\frac{\left[i(\frac{r'}{2s'})d(\frac{s'}{r'})-d\gamma\right]
	\:e^{i\gamma}}{\sqrt{\frac{(r'+s')^2(r'+s'+t')}{r's't'}}}\\
	\frac{\left[-i(\frac{r'}{2s'})d(\frac{s'}{r'})-d\gamma\right]
	\:e^{-i\gamma}}{\sqrt{\frac{(r'+s')^2(r'+s'+t')}{r's't'}}}~~~~~&
	-\frac{d\gamma}{1+\frac{r's'}{(r'+s')^2+r't'}}
	-\frac{d\theta}{1+\frac{t'}{r'+s'}}
	\end{array}\right)$}\;.
	\label{c1-a2}
	\xeea

\subsection*{Case 2: $r'=0$, $s'\neq 0$ and $t'\neq 0$}
In this case the eigenvectors are:
	\xbe
	|1\xkt=N_3\:\left(\begin{array}{c}
	0\\s'\\ \sqrt{s't'}\:e^{i\theta}
	\end{array}\right)\;,~~~~~
	|2,1\xkt=\left(\begin{array}{c}
	1\\0\\0
	\end{array}\right)\;,~~~~~
	|2,2\xkt=N_3\:\left(\begin{array}{c}
	0\\ \sqrt{s't'}\\ -s'\:e^{i\theta}
	\end{array}\right)\;,
	\label{eq-ve-c2}
	\xee
where $N_3:=[s'(s'+t')]^{-1/2}$. These equations together
with Eq.~(\ref{connection}) yield:
	\xbea
	A_1&=&\frac{-d\theta}{1+\frac{s'}{t'}}\;,
	\label{c2-a1}\\
	A_2&=&\left(\begin{array}{cc}
	0&0\\
	0&1
	\end{array}\right)\: (\frac{-d\theta}{1+\frac{t'}{s'}})\;.
	\label{c2-a2}
	\xeea
Note that in this case $A_2$ also leads to an Abelian
geometrical phase. 

\subsection*{Case 3: $s'=0$, $r'\neq 0$ and $t'\neq 0$}
In this case, the expressions for the eigenvectors and the
connection one-forms can be obtained from the
results of Case 2.  This is easily seen by the following
relationship between the Hamiltonian $H'_{(3)}$ for this
case and the Hamiltonian $H'_{(2)}$ for Case 2:
	\xbe
	H'_{(3)}=\left. T_1\:H'_{(2)}\:T^{-1}_1\right|_{r'\to s',~
	\theta\to -(\theta+\gamma)}\;,
	\label{h-trans}
	\xee
with
	\[ T_1:=\left(\begin{array}{ccc}
	0&1&0\\
	1&0&0\\
	0&0&1\end{array}\right)=T_1^{-1}\;.\]
In view of Eq.~(\ref{h-trans}), the eigenvectors
and the connection one-forms are related according to
	\xbea
	|n,a\xkt_{(3)}&=&\left. T_1|n,a\xkt_{(2)}\right|_{r'\to s',
	\theta\to -(\theta+\gamma)}\;,
	\label{eq-ve-c3}\\
	A_{n_{(3)}}&=&\left.A_{n_{(2)}}\right|_{r'\to s',
	\theta\to -(\theta+\gamma)}\;,
	\label{c3-a1-a2}
	\xeea
where the subscripts $(2)$ and $(3)$ correspond to the
cases $2$ and $3$.
\subsection*{Case 4: $t'=0$, $r'\neq 0$ and $s'\neq 0$}
The situation is analogous to the Case 3. Again one can
use the relations
	\xbe
	H'_{(4)}=\left. T_2\:H'_{(2)}\:T^{-1}_2\right|_{t'\to r',
	\theta\to -\gamma}\;,
	\label{h-trans'}
	\xee
with
	\[ T_2:=\left(\begin{array}{ccc}
	0&0&1\\
	0&1&0\\
	1&0&0\end{array}\right)=T_1^{-1}\;,\]
to read off  the expressions for the eigenvectors and
the connection one-forms from Eqs.(\ref{eq-ve-c2})-
(\ref{c2-a2}), namely
	\xbea
	|n,a\xkt_{(4)}&=&\left. T_2|n,a\xkt_{(2)}\right|_{t'\to r',
	\theta\to -\gamma}\;,
	\label{eq-ve-c4}\\
	A_{n_{(4)}}&=&\left.A_{n_{(2)}}\right|_{t'\to r',~
	\theta\to -\gamma}\;.
	\label{c4-a1-a2}
	\xeea

Clearly in all four cases the connection one-forms
depend only on the ratios of the parameters $r',s'$
and $t'$, as expected. The formulae for the connection
one-forms can be expressed in
terms of the parameters of the original Hamiltonian
$H$ using Eqs.~(\ref{primes=2}), (\ref{E1-E2-c2}),
(\ref{E1-E2-c3}), and (\ref{E1-E2-c4}) for cases
1, 2, 3, and 4, respectively.

\section{Application to Multipole Interactions}

The results of the preceding sections are clearly
applicable for arbitrary $3\times 3$ Hamiltonians.
In particular they can be used to compute non-Abelian
geometric phases of  spin $j=1$ systems with a
multipole interaction Hamiltonian of the form
	\xbe
	 H=Q^{i_1,\cdots i_n}J_{i_1}\cdots
	J_{i_n}\;,
	\label{h-multi}
	\xee
where $Q^{i_1,\cdots i_n}$ is symmetric in its
labels and $J_i$ are angular momentum operators.

The simplest example is the dipole Hamiltonian of
Eq.~(\ref{h1}). It is well-known that the eigenvalues
of this Hamiltonian are non-degenerate. This can also
be seen as a consequence of  the results of section~2.

Using the standard expressions for the matrix
representation of the angular momentum operators
for $j=1$, \cite{schiff}, the dipole
Hamiltonian (\ref{h1}) can be written in the form
(\ref{h4}) with\footnote{Here and in the following
calculations I have set $\hbar=1$ for convenience.}
	\xbe
	r=-t=R^3\;,~~~~~\xi=\kappa=\frac{R^1+i R^2}{
	\sqrt{2}}\;,~~~~~ \zeta=s=0\;.
%	\label{h1-matrix}
	\xee
Therefore either $\xi=\zeta=\kappa=0$ or $\zeta=0$
and $\xi\neq 0\neq\kappa$. In the former case the
Hamiltonian becomes diagonal with diagonal elements
being $R^3,~0,$ and $-R^3$, i.e., the eigenvalues
are not doubly degenerate. The latter case is a
particular example of Case~5 of section~2, for which
a doubly degenerate eigenvalue is again impossible.

The next simplest case is a quadrupole Hamiltonian
	\xbe
	H=Q^{ij}J_iJ_j\;.
	\label{h-quad}
	\xee
As I mentioned earlier, these Hamiltonians have been
studied for the fermionic systems in
Refs.~\cite{mead2,av1,av2}. It is shown in Ref.~\cite{av2}
that for the bosonic (integer spin) systems the Abelian
geometrical phases vanish in this case. The same argument
does not apply for the non-Abelian phases however. This
is quite easily seen by expressing the Hamiltonian
(\ref{h-quad}) in the form (\ref{h4}). This leads to:
	\[
	r=t=\frac{1}{2}(Q^{11}+Q^{22})+Q^{33},~~~~
	s=Q^{11}+Q^{22},~~~~
	\xi=-\kappa=\frac{Q^{13}+iQ^{23}}{\sqrt{2}},~~~~
	\zeta=\frac{1}{2}(Q^{11}-Q^{22})+iQ^{12}.
%	\label{h-quad-matrix}
	\]
Therefore either $\xi\neq 0$, in which case $\zeta\neq 0$,
for otherwise one has the same situation as in Case~5 of
section~2,  or $\xi=0$.  In the latter case one can
also assume that $\zeta\neq 0$, since $\xi=\zeta=0$
corresponds to the case where the eigenvectors of the
Hamiltonian are independent of the parameters and the
geometric phases are trivial. 

For $\zeta\neq 0\neq\xi$ and $\zeta\neq 0=\xi$
one has particular examples of Cases~1 and~3
of section~2, respectively. Therefore, in general
nontrivial geometric phases may exist.

An example of a quantum system with a quadrupole Hamiltonian (\ref{h-quad}) is the asymmetric rotor, for
which $Q^{12}=Q^{13}=Q^{23}=0$. In this case, one
has $\xi=\kappa=0$. Hence, the geometric phases
are trivial.

Another potentially interesting case is when both
dipole and quadrupole interactions are present, i.e.,
	\xbe
	H=R^iJ_i+Q^{jk}J_jJ_k\;.
	\label{h-dipole-quadrupole}
	\xee
For a spin $j=1$ system one can again express this
Hamiltonian in the form (\ref{h4}). This corresponds
to:
	\xbea
	r&=&R^3+\frac{1}{2}(Q^{11}+Q^{22})+Q^{33}\;,~~~~
	s\:=\:Q^{11}+Q^{22}\;,\xnn\\
	t&=&-R^3+\frac{1}{2}(Q^{11}+Q^{22})+Q^{33}\;,~~~~
	\xi\:=\:\frac{R^1+i R^2}{\sqrt{2}}+
	\frac{Q^{13}+iQ^{23}}{\sqrt{2}}\;,
%	\label{h12-matrix}
	\\
	\zeta&=&\frac{1}{2}(Q^{11}-Q^{22})+iQ^{12}\;,~~~~
	\kappa\:=\:\frac{R^1+i R^2}{\sqrt{2}}-		
	\frac{Q^{13}+iQ^{23}}{\sqrt{2}}\;.\xnn
	\xeea
Hence, in this case all possible cases of sections~2
may occur and nontrivial geometric phases
may be present. In particular, for an interaction
Hamiltonian of the form (\ref{h-dipole-quadrupole})
with the quadrupole part given by the asymmetric
rotor Hamiltonian, nontrivial  geometric phases
can exist.

\section{Conclusion}

In this article, I have solved the problem of  the adiabatic
non-Abelian geometrical phase for arbitrary quantum
systems with a three-dimensional Hilbert space.
This has been possible due to a very simple observation
that the eigenvalue problem for a three by three matrix
can be much easily handled if one knows that one of
the eigenvalues is doubly degenerate. 

The parameter space for all such systems can be
easily shown to be the projective space $\xC P^2$.
This is done using the well-known symmetry arguments.
An explicit construction of this space in  terms of the
matrix elements of the Hamiltonian however involves
solving the eigenvalue problem.  Therefore it is not
an easy task. The indirect but efficient method
of solving the eigenvalue problem which I have
employed in this article, also leads to an explicit
construction of the parameter space.

The results of this article may be applied for arbitrary
spin $j=1$ systems. In particular, I have discussed
the dipole, quadrupole, and a combination of a dipole
and quadrupole Hamiltonians. A simple example of
a quadrupole Hamiltonian is that of the asymmetric
rotor. The geometric phases for this system turn out
to be trivial. The addition of an appropriate dipole term
to the Hamiltonian of the asymmetric rotor, however,
does lead to nontrivial geometric phases.

\section*{Acknowledgements}
I would like to thank Bahman Darian for his patiently
listening to my arguments, and  wish to acknowledge the
support of the Killam Foundation of Canada.

\end{document}